\pdfoutput=1
\documentclass[11pt]{article}

\usepackage[final]{acl}

\usepackage{times}
\usepackage{latexsym}
\usepackage{titlesec}

\usepackage[T1]{fontenc}
\usepackage[utf8]{inputenc}

\usepackage{microtype}

\usepackage{inconsolata}

\usepackage{graphicx}

\usepackage{times}
\usepackage{latexsym}
\usepackage{kotex}
\usepackage{todonotes}
\usepackage{arydshln}
\usepackage{tabularx}
\usepackage{booktabs}
\usepackage{graphicx}
\usepackage{amsmath}
\usepackage{makecell}
\usepackage{array}
\usepackage{tabularx}
\usepackage{multirow}
\usepackage{siunitx}
\usepackage{adjustbox}
\usepackage{rotating}
\usepackage[utf8]{inputenc}
\usepackage{caption}
\usepackage{hyperref}

\usepackage{booktabs}
\usepackage{multirow}
\usepackage{colortbl}
\usepackage{pgfmath}
\usepackage{nicematrix}
\usepackage{tcolorbox}

\title{Improving Korean-English Cross-Lingual Retrieval: A Data-Centric Study of Language Composition and Model Merging}
\author{
Youngjoon Jang$^{1}$, Junyoung Son$^{1}$, Taemin Lee$^{1}$, Seongtae Hong$^{1}$, Hyeonseok Moon$^{1}$ \\
\textbf{Seungyoon Lee$^{1}$, Andrew Matteson$^{2}$, Heuiseok Lim$^{1\dagger}$} \\
\\
$^{1}$Korea University \quad $^{2}$Uracle \\
\texttt{\{dew1701, s0ny, taeminlee, ghdchlwls123, glee889, dltmddbs100, limhseok\}@korea.ac.kr} \\
\texttt{amatteson@uracle.co.kr}
}

\newcommand\blfootnote[1]{%
  \begingroup
  \renewcommand\thefootnote{}\footnote{#1}%
  \addtocounter{footnote}{-1}%
  \endgroup
}

\begin{document}
\maketitle
\begin{abstract}
% ARR Oct 원본
% With the increasing utilization of multilingual text information, Cross-Lingual Information Retrieval~(CLIR) has become a crucial research area. However, the impact of training data composition on CLIR and Mono-Lingual Information Retrieval~(Mono-IR) performance remains underexplored. To investigate this data-centric aspect, we construct linguistically parallel Korean-English datasets and train multilingual retrieval models with various language combinations. Our experiments reveal that the language composition of training data significantly influences IR performance, exhibiting important inter-lingual correlations: Using specific language pairs improves CLIR performance, while reducing Mono-IR performance. Our work demonstrates that simple weight-averaged model merging can effectively mitigate this trade-off, achieving strong CLIR results while preserving Mono-IR capabilities. Our findings highlight the effects of linguistic configuration of training data on both CLIR and Mono-IR, and present model merging as a viable strategy to optimize performance across these tasks.

With the increasing utilization of multilingual text information, Cross-Lingual Information Retrieval~(CLIR) has become a crucial research area. However, the impact of training data composition on CLIR and Mono-Lingual Information Retrieval~(Mono-IR) performance remains underexplored.
In this work, we conduct a data-centric study on Korean--English CLIR by systematically varying the language composition of the dataset for dense retriever fine-tuning.
Using linguistically parallel training sets that support all cross-lingual permutations, we analyze how different compositions shape retrieval behavior across multiple embedding models and evaluation benchmarks.
Our results reveal a consistent trade-off: configurations that substantially improve CLIR can simultaneously degrade Mono-IR, indicating that retrieval gains are coupled through inter-lingual correlations induced during training.
We further identify which components of the training data are primarily responsible for the observed gains and regressions, providing actionable guidance for dataset construction.
Finally, as a lightweight practical option, we show that a simple post-hoc weight averaging with the base model can partially recover Mono-IR performance while retaining most of the CLIR improvements.

\end{abstract}

\section{Introduction}
Information Retrieval (IR) aims to retrieve documents that are relevant to a user query, and it has become a core component of modern Retrieval-Augmented Generation (RAG) and agentic systems~\citep{lewis2020retrieval, gao2023retrieval, singh2025agentic}. Although the significance of IR is widely recognized, the majority of research has centered on monolingual settings~\citep{zhao2024dense}. This focus often overlooks Cross-Lingual Information Retrieval (CLIR) scenarios, where the languages of the queries and documents differ.
\blfootnote{$^\dagger$Corresponding author.}
Previous CLIR studies introduce various training data compositions to enhance cross-lingual performance. This includes pairing queries and passages in different languages and diversifying the languages of query–passage pairs~\citep{goswami2021cross, zhang2022mind, gao2001improving, bhattacharya2016query, saleh2020document}. 
Despite these efforts, there has been scarce discussion on how different training data compositions affect the performances of Mono-IR and CLIR.

In this work, we present a data-centric study of Korean--English CLIR that systematically examines the role of language composition in training data. We first construct a synthetic Korean--English parallel training dataset that supports controlled variations of language assignments within training triples. Using this dataset, we fine-tune dense retrievers under different compositions and analyze their effects on both Mono-IR and CLIR across multiple models and evaluation benchmarks. Our experiments reveal a consistent trade-off: certain language pairings in training substantially improve CLIR in a given direction while degrading Mono-IR, indicating that training data composition can induce coupled changes in retrieval behavior.

Beyond documenting this trade-off, we identify which aspects of composition most strongly influence the retrieval performance, providing guidance for constructing CLIR training data. Finally, as a lightweight practical baseline, we report that simple weight-averaged model merging~\citep{matena2022merging} with the base model can partially reduce the Mono-IR drop while retaining much of the CLIR improvement.

\section{Related Work}

Information Retrieval (IR) aims to retrieve documents that are relevant to a given query~\citep{singhal2001modern, zhu2023large, li2025matching}. 
Recent retrieval methods include dense retrieval~\citep{karpukhin2020dense, xiong2020approximate, wang2024multilingual, zhang2024mgte, chen-etal-2024-m3}, sparse retrieval~\citep{dai2020context, formal2021splade, formal2021spladev2}, and multi-vector retrieval~\citep{khattab2020colbert, liu2023understanding}. 
Most of these approaches have been developed and evaluated primarily in monolingual settings, where the query and document collection are written in the same language.

Cross-Lingual Information Retrieval (CLIR) extends this setting by retrieving relevant documents written in a language different from that of the query. 
Recent studies have improved CLIR through multilingual representation learning, cross-lingual pretraining, translation-based supervision, and knowledge transfer across languages~\citep{goswami2021cross, yu2021cross, litschko2022cross, huang2023improving, nguyen2025improving}. 
While these studies mainly focus on improving cross-lingual retrieval accuracy, the role of training data composition itself remains less explored.

In this work, we conduct a data-centric analysis of Korean--English CLIR by isolating the language composition of training triples. 
Specifically, we systematically vary the languages of queries, positives, and negatives while keeping the underlying semantics parallel across variants. 
This controlled setup allows us to analyze how different language compositions affect both CLIR and monolingual retrieval behavior, revealing a trade-off between cross-lingual gains and monolingual degradation.

\section{Synthetic Data Generation}
Our study aims to analyze the impact of language composition on IR performance. As recent embedding models utilize a diverse range of datasets, employing open-source data might risk overlapping with previously trained data. Furthermore, there is a scarcity of linguistically parallel query-positive (a passage providing answer to query) data. To address these issues and more robustly assess the effect of language composition, we build a high-quality training dataset that is linguistically parallel and covers a diverse range of domains.
\paragraph{KDC Sampling}
To ensure broad domain coverage and overcome potential LLM domain limitations, we incorporate the Korean Decimal Classification (KDC)~\citep{oh2020korean}. KDC is a Korean adaptation of the Dewey Decimal Classification (DDC)~\citep{scott1998dewey}, a comprehensive library classification system. KDC has a hierarchical structure, where each digit represents a specific subject level: the first digit signifies a broad subject area, the second a subcategory, and the third a more specific category. For example, in the code 761, the 7 stands for the main class 'Language', the 6 for the subclass 'French language', and the 1 for the specific topic 'Phonology and writing systems'. We use all levels of KDC, which consists of 910 topics. 

\paragraph{Task Brainstorming \& Triple Generation}
Following the synthetic data construction procedure of E5-Mistral~\cite{E5_Mistral}, we generate training triples in two steps.
For each sampled KDC category, we first obtain a set of task descriptions, which are then used as prompts to produce \{query, positive, hard\_negative\} triples.
We deduplicate the resulting queries and passages, and mine six hard negatives per query to strengthen training signals.
Finally, we construct six Korean--English cross-lingual variants of the triples to cover all cross-lingual permutations.
Additional details on dataset generation, dataset statistics, and human quality assessment are provided in Appendix~\ref{sec:data_details}.

% % 100 곱하고, AVG는 유지한 버전

\definecolor{metacolorab}{HTML}{F0F0F0}      % very-light gray
% Pastel Red Tones
\definecolor{strongred}{HTML}{F4A7A3}     % Light Coral (adjusting for 'strong' pastel)
\definecolor{mediumred}{HTML}{F8C8C5}     % Lighter Coral
\definecolor{lightred}{HTML}{FCEAE9}      % Very Light Coral/Pinkish
% Pastel Green Tones
\definecolor{stronggreen}{HTML}{A2D9B8}    % Light Mint (adjusting for 'strong' pastel)
\definecolor{mediumgreen}{HTML}{C5E8D4}    % Lighter Mint
\definecolor{lightgreen}{HTML}{E9F7F0}     % Very Light Mint

% 성태성태성태
% \definecolor{stronggreen}{HTML}{A2DFF7}    % 중심색 – 맑고 세련된 시안 블루
% \definecolor{mediumgreen}{HTML}{D0F0FB}    % 중심보다 밝고 부드러운 연파랑
% \definecolor{lightgreen}{HTML}{F2FAFD}     % 거의 흰색에 가까운 연청색 배경

\begin{table*}[!ht]
\centering
% \small
\resizebox{1 \linewidth}{!}{%

\begin{tabular}{c|cc|cc!{\vrule width 1pt}cc|c||cc|cc!{\vrule width 1pt}cc|c}
\toprule[1.2pt]
\multirow{4}{*}{\textbf{Data}} &
\multicolumn{7}{c||}{\textbf{Cross-Lingual}} &
\multicolumn{7}{c}{\textbf{Mono-Lingual}} \\
\cmidrule(lr){2-8}\cmidrule(lr){9-15}
& \multicolumn{2}{c|}{\textbf{Belebele}} & \multicolumn{2}{c!{\vrule width 1pt}}{\textbf{StrategyQA}} & \multicolumn{2}{c|}{\textbf{AVG}} & \multirow{2}{*}{\textbf{OVR}} &
\multicolumn{2}{c|}{\textbf{Belebele}} & \multicolumn{2}{c!{\vrule width 1pt}}{\textbf{StrategyQA}} & \multicolumn{2}{c|}{\textbf{AVG}} & \multirow{2}{*}{\textbf{OVR}} \\
\cmidrule(lr){2-3}\cmidrule(lr){4-5}\cmidrule(lr){6-7} \cmidrule(lr){9-10}\cmidrule(lr){11-12}\cmidrule(lr){13-14}
\scalebox{1}{Q-P-N} & en-ko & ko-en & en-ko & ko-en & en-ko & ko-en & & ko-ko & en-en & ko-ko & en-en & ko-ko & en-en & \\

\midrule[1.2pt]
\multicolumn{15}{c}{\textit{\textbf{bge-m3}}}\\
\midrule[1.2pt]
\rowcolor{metacolorab} base & 90.37 & 88.36 & 81.24 & 71.65 & 85.81 & 80.01 & 82.91 & 93.16 & 95.55 & 79.41 & 84.42 & 86.29 & 89.99 & \textbf{88.14} \\
kokoen & \cellcolor{lightred} 90.32 & \cellcolor{lightred} 87.95 & \cellcolor{lightgreen} 81.64 & \cellcolor{lightred} 71.39 & \cellcolor{lightgreen} 85.98 & \cellcolor{lightred} 79.67 & \cellcolor{lightred} 82.83 & \cellcolor{lightred} 92.80 & \cellcolor{lightred} 95.16 & \cellcolor{lightgreen} 79.68 & \cellcolor{lightred} 84.29 & \cellcolor{lightred} \underline{86.24} & \cellcolor{lightred} 89.73 & \cellcolor{lightred} 87.98 \\
koenko & \cellcolor{lightred} 89.32 & \cellcolor{lightgreen} 89.08 & \cellcolor{mediumred} 79.44 & \cellcolor{mediumgreen} 72.92 & \cellcolor{mediumred} 84.38 & \cellcolor{lightgreen} \underline{81.00} & \cellcolor{lightred} 82.69 & \cellcolor{mediumred} 91.77 & \cellcolor{lightred} 94.64 & \cellcolor{mediumred} 78.06 & \cellcolor{lightred} 83.92 & \cellcolor{mediumred} 84.92 & \cellcolor{lightred} 89.28 & \cellcolor{strongred} 87.10 \\
koenen & \cellcolor{mediumgreen} 91.46 & \cellcolor{mediumgreen} 89.65 & \cellcolor{lightgreen} 81.69 & \cellcolor{mediumgreen} 72.86 & \cellcolor{lightgreen} \underline{86.58} & \cellcolor{mediumgreen} \textbf{81.26} & \cellcolor{stronggreen} \textbf{83.92} & \cellcolor{lightred} 92.65 & \cellcolor{lightred} 95.03 & \cellcolor{lightred} 79.13 & \cellcolor{lightred} 84.06 & \cellcolor{lightred} 85.89 & \cellcolor{lightred} 89.55 & \cellcolor{lightred} 87.72 \\
enenko & \cellcolor{mediumred} 89.09 & \cellcolor{lightgreen} 88.58 & \cellcolor{lightred} 80.76 & \cellcolor{lightgreen} 71.68 & \cellcolor{lightred} 84.93 & \cellcolor{lightgreen} 80.13 & \cellcolor{lightred} 82.53 & \cellcolor{lightgreen} 93.19 & \cellcolor{lightred} 95.38 & \cellcolor{lightred} 79.34 & \cellcolor{lightgreen} 84.43 & \cellcolor{lightred} \textbf{86.27} & \cellcolor{lightred} \underline{89.91} & \cellcolor{lightred} 88.09 \\
enkoen & \cellcolor{mediumgreen} 91.59 & \cellcolor{lightred} 88.29 & \cellcolor{lightgreen} 81.68 & \cellcolor{lightgreen} 71.89 & \cellcolor{lightgreen} \textbf{86.64} & \cellcolor{lightgreen} 80.09 & \cellcolor{lightgreen} 83.36 & \cellcolor{mediumred} 91.89 & \cellcolor{lightred} 94.78 & \cellcolor{lightred} 78.99 & \cellcolor{lightred} 83.92 & \cellcolor{lightred} 85.44 & \cellcolor{lightred} 89.35 & \cellcolor{mediumred} 87.40 \\
enkoko & \cellcolor{lightgreen} 91.24 & \cellcolor{lightgreen} 89.00 & \cellcolor{lightgreen} 81.81 & \cellcolor{mediumgreen} 72.73 & \cellcolor{lightgreen} 86.53 & \cellcolor{lightgreen} 80.87 & \cellcolor{mediumgreen} \underline{83.70} & \cellcolor{lightred} 92.29 & \cellcolor{lightred} 94.83 & \cellcolor{lightred} 79.31 & \cellcolor{lightred} 84.24 & \cellcolor{lightred} 85.80 & \cellcolor{lightred} 89.54 & \cellcolor{lightred} 87.67 \\
\midrule[1.2pt]
\multicolumn{15}{c}{\textit{\textbf{multilingual-e5-large (mE5-large)}}}\\
\midrule[1.2pt]
\rowcolor{metacolorab} base & 92.03 & 86.45 & 82.04 & 67.30 & 87.04 & 76.88 & 81.96 & 94.50 & 96.50 & 80.35 & 84.20 & 87.43 & 90.35 & \cellcolor{lightred} \textbf{88.89} \\
kokoen & \cellcolor{lightgreen} 92.82 & \cellcolor{lightred} 86.40 & \cellcolor{lightgreen} 82.63 & \cellcolor{stronggreen} 70.72 & \cellcolor{lightgreen} 87.73 & \cellcolor{mediumgreen} 78.56 & \cellcolor{stronggreen} 83.14 & \cellcolor{mediumred} 92.87 & \cellcolor{lightred} 95.84 & \cellcolor{lightred} 80.13 & \cellcolor{lightgreen} 84.71 & \cellcolor{lightred} 86.50 & \cellcolor{lightred} \underline{90.28} & \cellcolor{mediumred} 88.39 \\
koenko & \cellcolor{strongred} 81.02 & \cellcolor{stronggreen} 89.43 & \cellcolor{strongred} 68.20 & \cellcolor{stronggreen} 73.33 & \cellcolor{strongred} 74.61 & \cellcolor{stronggreen} \textbf{81.38} & \cellcolor{strongred} 78.00 & \cellcolor{strongred} 84.66 & \cellcolor{mediumred} 94.60 & \cellcolor{strongred} 71.11 & \cellcolor{lightred} 84.08 & \cellcolor{strongred} 77.89 & \cellcolor{mediumred} 89.34 & \cellcolor{strongred} 83.61 \\
koenen & \cellcolor{mediumgreen} 93.34 & \cellcolor{stronggreen} 89.27 & \cellcolor{lightgreen} 82.63 & \cellcolor{stronggreen} 72.98 & \cellcolor{lightgreen} \underline{87.99} & \cellcolor{stronggreen} \underline{81.13} & \cellcolor{stronggreen} \textbf{84.56} & \cellcolor{mediumred} 92.12 & \cellcolor{mediumred} 95.12 & \cellcolor{lightred} 79.89 & \cellcolor{lightred} 84.13 & \cellcolor{mediumred} 86.01 & \cellcolor{lightred} 89.63 & \cellcolor{strongred} 87.82 \\
% enenen & \cellcolor{lightred} 91.84 & \cellcolor{lightred} 86.28 & \cellcolor{lightgreen} 82.55 & \cellcolor{mediumgreen} 70.11 & \cellcolor{lightgreen} 87.20 & \cellcolor{mediumgreen} 78.20 & \cellcolor{mediumgreen} 82.70 & \cellcolor{mediumred} 92.91 & \cellcolor{mediumred} 95.24 & \cellcolor{lightgreen} 80.39 & \cellcolor{lightgreen} 84.62 & \cellcolor{lightred} \textbf{86.65} & \cellcolor{lightred} 89.93 & \cellcolor{mediumred} 88.29 \\
enenko & \cellcolor{mediumred} 90.65 & \cellcolor{mediumgreen} 87.66 & \cellcolor{mediumred} 80.46 & \cellcolor{mediumgreen} 70.21 & \cellcolor{mediumred} 85.56 & \cellcolor{mediumgreen} 78.94 & \cellcolor{lightgreen} 82.25 & \cellcolor{mediumred} 92.96 & \cellcolor{mediumred} 95.47 & \cellcolor{lightred} 80.14 & \cellcolor{lightgreen} 84.94 & \cellcolor{lightred} \underline{86.55} & \cellcolor{lightred} 90.21 & \cellcolor{mediumred} 88.38 \\
enkoen & \cellcolor{lightgreen} 93.05 & \cellcolor{strongred} 74.77 & \cellcolor{lightgreen} 82.11 & \cellcolor{strongred} 58.88 & \cellcolor{lightgreen} 87.58 & \cellcolor{strongred} 66.83 & \cellcolor{strongred} 77.20 & \cellcolor{strongred} 91.22 & \cellcolor{strongred} 89.94 & \cellcolor{mediumred} 79.33 & \cellcolor{strongred} 79.98 & \cellcolor{mediumred} 85.28 & \cellcolor{strongred} 84.96 & \cellcolor{strongred} 85.12 \\
enkoko & \cellcolor{mediumgreen} 93.17 & \cellcolor{mediumgreen} 88.95 & \cellcolor{mediumgreen} 82.91 & \cellcolor{stronggreen} 71.92 & \cellcolor{mediumgreen} \textbf{88.04} & \cellcolor{stronggreen} 80.44 & \cellcolor{stronggreen} \underline{84.24} & \cellcolor{strongred} 91.46 & \cellcolor{lightred} 95.63 & \cellcolor{lightred} 80.06 & \cellcolor{lightgreen} 84.45 & \cellcolor{mediumred} 85.76 & \cellcolor{lightred} 90.04 & \cellcolor{mediumred} 87.90 \\
\midrule[1.2pt]
\multicolumn{15}{c}{\textit{\textbf{multilingual-e5-base (mE5-base)}}}\\
\midrule[1.2pt]
\rowcolor{metacolorab} base & 82.28 & 77.52 & 77.94 & 54.44 & 80.11 & 65.98 & 73.05 & 92.87 & 95.79 & 76.36 & 83.21 & 84.62 & 89.50 & \textbf{87.06} \\
% kokoko & \cellcolor{mediumgreen} 83.97 & \cellcolor{mediumgreen} 79.75 & \cellcolor{mediumgreen} 79.15 & \cellcolor{stronggreen} 58.23 & \cellcolor{mediumgreen} 81.56 & \cellcolor{stronggreen} 68.99 & \cellcolor{stronggreen} 75.28 & \cellcolor{mediumred} 90.81 & \cellcolor{lightred} 95.33 & \cellcolor{lightred} 75.43 & \cellcolor{lightgreen} 83.89 & \cellcolor{mediumred} 83.12 & \cellcolor{lightgreen} 89.61 & \cellcolor{mediumred} 86.37 \\
kokoen & \cellcolor{stronggreen} 85.26 & \cellcolor{mediumgreen} 79.34 & \cellcolor{mediumgreen} 79.19 & \cellcolor{stronggreen} 57.89 & \cellcolor{mediumgreen} 82.23 & \cellcolor{mediumgreen} 68.62 & \cellcolor{stronggreen} 75.42 & \cellcolor{mediumred} 90.55 & \cellcolor{lightgreen} 95.81 & \cellcolor{lightred} 75.80 & \cellcolor{lightgreen} 84.01 & \cellcolor{mediumred} 83.18 & \cellcolor{lightgreen} \textbf{89.91} & \cellcolor{mediumred} 86.54 \\
koenko & \cellcolor{strongred} 75.30 & \cellcolor{stronggreen} 83.35 & \cellcolor{strongred} 60.05 & \cellcolor{stronggreen} 62.84 & \cellcolor{strongred} 67.68 & \cellcolor{stronggreen} \underline{73.10} & \cellcolor{strongred} 70.39 & \cellcolor{strongred} 83.31 & \cellcolor{mediumred} 93.75 & \cellcolor{strongred} 62.48 & \cellcolor{lightred} 82.68 & \cellcolor{strongred} 72.90 & \cellcolor{mediumred} 88.22 & \cellcolor{strongred} 80.56 \\
koenen & \cellcolor{stronggreen} 89.02 & \cellcolor{stronggreen} 83.29 & \cellcolor{mediumgreen} 80.06 & \cellcolor{stronggreen} 63.23 & \cellcolor{stronggreen} 84.54 & \cellcolor{stronggreen} \textbf{73.26} & \cellcolor{stronggreen} \textbf{78.90} & \cellcolor{strongred} 89.61 & \cellcolor{mediumred} 94.17 & \cellcolor{mediumred} 74.71 & \cellcolor{lightgreen} 83.65 & \cellcolor{mediumred} 82.16 & \cellcolor{lightred} 88.91 & \cellcolor{strongred} 85.54 \\
% enenen & \cellcolor{stronggreen} 86.64 & \cellcolor{mediumgreen} 79.68 & \cellcolor{mediumgreen} 79.32 & \cellcolor{stronggreen} 57.90 & \cellcolor{stronggreen} 82.98 & \cellcolor{mediumgreen} 68.79 & \cellcolor{stronggreen} 75.89 & \cellcolor{mediumred} 91.01 & \cellcolor{lightred} 95.35 & \cellcolor{lightred} 75.43 & \cellcolor{lightgreen} 84.09 & \cellcolor{mediumred} \textbf{83.22} & \cellcolor{lightgreen} \underline{89.72} & \cellcolor{mediumred} 86.47 \\
enenko & \cellcolor{stronggreen} 85.17 & \cellcolor{stronggreen} 80.31 & \cellcolor{lightred} 76.93 & \cellcolor{stronggreen} 58.59 & \cellcolor{lightgreen} 81.05 & \cellcolor{stronggreen} 69.45 & \cellcolor{stronggreen} 75.25 & \cellcolor{mediumred} 90.92 & \cellcolor{lightred} 95.22 & \cellcolor{lightred} 75.48 & \cellcolor{lightgreen} 84.20 & \cellcolor{mediumred} \underline{83.20} & \cellcolor{lightgreen} 89.71 & \cellcolor{mediumred} 86.46 \\
enkoen & \cellcolor{stronggreen} 88.94 & \cellcolor{lightgreen} 77.93 & \cellcolor{mediumgreen} 80.29 & \cellcolor{lightgreen} 55.28 & \cellcolor{stronggreen} \underline{84.62} & \cellcolor{lightgreen} 66.61 & \cellcolor{stronggreen} 75.61 & \cellcolor{strongred} 87.37 & \cellcolor{mediumred} 94.74 & \cellcolor{mediumred} 73.90 & \cellcolor{lightred} 82.35 & \cellcolor{strongred} 80.64 & \cellcolor{lightred} 88.55 & \cellcolor{strongred} 84.59 \\
enkoko & \cellcolor{stronggreen} 89.43 & \cellcolor{stronggreen} 82.50 & \cellcolor{stronggreen} 80.60 & \cellcolor{stronggreen} 61.83 & \cellcolor{stronggreen} \textbf{85.02} & \cellcolor{stronggreen} 72.17 & \cellcolor{stronggreen} \underline{78.59} & \cellcolor{strongred} 88.10 & \cellcolor{lightred} 94.79 & \cellcolor{mediumred} 73.76 & \cellcolor{lightgreen} 83.90 & \cellcolor{strongred} 80.93 & \cellcolor{lightred} 89.35 & \cellcolor{strongred} 85.14 \\
\midrule[1.2pt]
\multicolumn{15}{c}{\textit{\textbf{gte-multilingual-base (mgte-base)}}}\\
\midrule[1.2pt]
\rowcolor{metacolorab} base & 88.06 & 85.66 & 80.54 & 64.38 & 84.30 & 75.02 & 79.66 & 87.96 & 94.63 & 75.12 & 84.26 & 81.54 & 89.45 & 85.49 \\
% kokoko & \cellcolor{mediumgreen} 90.29 & \cellcolor{lightred} 85.42 & \cellcolor{lightred} 80.21 & \cellcolor{mediumred} 62.93 & \cellcolor{lightgreen} 85.25 & \cellcolor{lightred} 74.18 & \cellcolor{lightgreen} 79.71 & \cellcolor{stronggreen} 91.74 & \cellcolor{lightgreen} 95.60 & \cellcolor{mediumgreen} 77.19 & \cellcolor{lightgreen} 84.42 & \cellcolor{mediumgreen} \textbf{84.47} & \cellcolor{lightgreen} 90.01 & \cellcolor{stronggreen} \textbf{87.24} \\
kokoen & \cellcolor{mediumgreen} 90.32 & \cellcolor{strongred} 82.31 & \cellcolor{lightgreen} 80.97 & \cellcolor{mediumred} 63.33 & \cellcolor{mediumgreen} 85.65 & \cellcolor{mediumred} 72.82 & \cellcolor{lightred} 79.23 & \cellcolor{stronggreen} 91.57 & \cellcolor{lightred} 94.21 & \cellcolor{mediumgreen} 76.92 & \cellcolor{lightgreen} 84.28 & \cellcolor{mediumgreen} \underline{84.25} & \cellcolor{lightred} 89.25 & \cellcolor{stronggreen} \underline{86.75} \\
koenko & \cellcolor{strongred} 80.19 & \cellcolor{lightgreen} 86.43 & \cellcolor{strongred} 76.19 & \cellcolor{lightgreen} 64.91 & \cellcolor{strongred} 78.19 & \cellcolor{lightgreen} 75.67 & \cellcolor{strongred} 76.93 & \cellcolor{strongred} 82.62 & \cellcolor{lightgreen} 94.86 & \cellcolor{strongred} 68.53 & \cellcolor{lightgreen} 84.63 & \cellcolor{strongred} 75.58 & \cellcolor{lightgreen} 89.75 & \cellcolor{strongred} 82.66 \\
koenen & \cellcolor{mediumgreen} 89.95 & \cellcolor{lightgreen} 86.56 & \cellcolor{lightgreen} 81.22 & \cellcolor{lightgreen} 65.10 & \cellcolor{mediumgreen} 85.59 & \cellcolor{lightgreen} \underline{75.83} & \cellcolor{stronggreen} 80.71 & \cellcolor{mediumgreen} 89.21 & \cellcolor{lightgreen} 94.71 & \cellcolor{lightgreen} 75.44 & \cellcolor{lightgreen} 84.48 & \cellcolor{lightgreen} 82.33 & \cellcolor{lightgreen} 89.60 & \cellcolor{lightgreen} 85.96 \\
% enenen & \cellcolor{mediumgreen} 90.28 & \cellcolor{mediumgreen} 87.48 & \cellcolor{lightgreen} 81.05 & \cellcolor{lightgreen} 64.39 & \cellcolor{mediumgreen} 85.67 & \cellcolor{lightgreen} \textbf{75.94} & \cellcolor{stronggreen} \underline{80.80} & \cellcolor{mediumgreen} 90.54 & \cellcolor{lightgreen} 95.42 & \cellcolor{lightgreen} 75.95 & \cellcolor{lightgreen} 84.61 & \cellcolor{mediumgreen} 83.25 & \cellcolor{lightgreen} \underline{90.02} & \cellcolor{stronggreen} 86.63 \\
enenko & \cellcolor{mediumred} 86.67 & \cellcolor{mediumgreen} 87.09 & \cellcolor{mediumred} 79.13 & \cellcolor{lightgreen} 64.53 & \cellcolor{mediumred} 82.90 & \cellcolor{lightgreen} 75.81 & \cellcolor{lightred} 79.36 & \cellcolor{mediumgreen} 89.38 & \cellcolor{mediumgreen} 95.69 & \cellcolor{lightred} 75.08 & \cellcolor{lightgreen} 84.47 & \cellcolor{lightgreen} 82.23 & \cellcolor{lightgreen} \textbf{90.08} & \cellcolor{mediumgreen} 86.16 \\
enkoen & \cellcolor{mediumgreen} 90.31 & \cellcolor{strongred} 77.49 & \cellcolor{lightgreen} 81.26 & \cellcolor{lightred} 64.16 & \cellcolor{mediumgreen} \underline{85.79} & \cellcolor{strongred} 70.83 & \cellcolor{strongred} 78.31 & \cellcolor{mediumgreen} 89.81 & \cellcolor{strongred} 89.91 & \cellcolor{lightgreen} 75.49 & \cellcolor{lightred} 84.00 & \cellcolor{mediumgreen} 82.65 & \cellcolor{mediumred} 86.96 & \cellcolor{mediumred} 84.80 \\
enkoko & \cellcolor{stronggreen} 90.64 & \cellcolor{lightgreen} 86.12 & \cellcolor{mediumgreen} 81.56 & \cellcolor{lightgreen} 64.93 & \cellcolor{mediumgreen} \textbf{86.10} & \cellcolor{lightgreen} 75.53 & \cellcolor{stronggreen} \textbf{80.81} & \cellcolor{mediumgreen} 89.66 & \cellcolor{lightgreen} 95.12 & \cellcolor{lightgreen} 75.71 & \cellcolor{lightgreen} 84.81 & \cellcolor{mediumgreen} 82.69 & \cellcolor{lightgreen} 89.97 & \cellcolor{mediumgreen} 86.33 \\
\bottomrule[1.2pt]
\end{tabular}%
}
\caption{CLIR (Left) and Mono-IR (Right) performance. Data notation: languages of query, positive, negatives (e.g., kokoen for Korean query, Korean positive and English negatives). Task notation: $Q_{L}-P_{L}$ (e.g., en-ko for English query, Korean documents). 
Highest scores in \textbf{bold}, second highest \underline{underlined}. 
Cell colors indicate changes relative to the corresponding base model, with green/red denoting improvement/degradation. OVR denotes the micro-average over the two language directions in each block:
\textit{en-ko}/\textit{ko-en} for CLIR and \textit{ko-ko}/\textit{en-en} for Mono-IR.}
% OVR refers to the micro average of both \textit{en-ko}, \textit{ko-en} tasks.}
\label{tab:main1}
\end{table*}

\section{Experimental Setup}
\paragraph{Training Details}
We fine-tune four base models: bge-m3~\citep{chen-etal-2024-m3}, multilingual-e5-large, multilingual-e5-base~\citep{wang2024multilingual}, and gte-multilingual-base~\citep{zhang2024mgte} using each of the language combination datasets constructed in the previous section. The training employs GISTEmbedLoss~\citep{solatorio2024gistembed}, a contrastive loss function, reducing noise from random in-batch negatives. Comprehensive training parameters and environment details are provided in Appendix~\ref{appen_exp_setup}.
\paragraph{Evaluation Datasets \& Metrics}
% To assess the retrieval performance of the trained embedding models, we utilize two Korean-English parallel retrieval datasets: Belebele~\citep{bandarkar2023belebele} and StrategyQA~\citep{geva2021did}. 
Since there are few retrieval benchmarks covering both Korean and English, we employ two retrieval benchmarks: Belebele~\cite{bandarkar2023belebele} and StrategyQA~\citep{geva2021did}. Both benchmarks are included in the official retrieval tasks of Massive Multilingual Text Embedding Benchmark(MMTEB)~\cite{enevoldsen2025mmteb}, which are widely adopted in contemporary retrieval works. Details about our evaluation benchmark are stated in Appendix~\ref{sec:appen_benchmark}.
% Both datasets feature parallel queries and documents in Korean and English.
%, enabling a comprehensive evaluation of both CLIR and Mono-Lingual IR performance
We report nDCG@10~\citep{wang2013theoretical} as our primary evaluation metric, following the convention of MMTEB.
\paragraph{Data Notation}
We use specific notations to describe the language composition of our training data and evaluation task. The training data language notation $\mathcal{D}(l_1,l_2,l_3)$ specifies the languages of the query, positive, and negatives sequentially. For example, $\mathcal{D}(ko,en,ko)$ indicates a Korean query, an English positive, and Korean negatives dataset. Task notations $\mathcal{T}(l_1-l_2)$ indicate the languages of the query and the document pool respectively. For instance, $\mathcal{T}(en- ko)$ refers to a task for retrieving relevant documents from a Korean document pool given an English query. Additionally, to simplify exposition in the Analysis section, we denote the query language by $Q_{L}$, the positive language by $P_{L}$, and the language of the negatives by $N_{L}$.

\section{Experiments and Analysis}
\label{sec:analysis}
\subsection{Cross-Lingual Retrieval Results}
On the left side of Table~\ref{tab:main1}, we present a comprehensive summary of CLIR performance for the base models and all six training data combinations.

% \paragraph{Enhancement by Cross-Lingual Query-Positive}
% Training with query-positive pairs that mirror the language direction of a given CLIR task consistently improves performance. For example, in the $\mathcal{T}(en\texttt{-}ko)$ task, the bge-m3 model trained with $\mathcal{D}(enkoen)$ achieves a higher AVG score (86.64) than when trained with $\mathcal{D}(enenko)$ (84.93). Similarly, for the mE5-base model on the $\mathcal{T}(ko\texttt{-}en)$ task, training with $\mathcal{D}(koenen)$ (73.26) or $\mathcal{D}(koenko)$ (73.10) outperforms the base model (65.98) or training with $\mathcal{D}(kokoen)$ (68.62).
% These results suggest that cross-lingual query-positive pairs encourage the model to learn semantic similarities from different languages, thereby enhancing CLIR performance. 
\paragraph{Enhancement by task-aligned Query–Positive}
For a given CLIR language direction $\mathcal{T}(Q_L- P_L)$, training on query–positive pairs that match this direction improves performance on the specific task. 
For example, in the $\mathcal{T}(en-ko)$ task, the bge-m3 model trained with $\mathcal{D}(en,ko,en)$ achieves a higher AVG score (86.64) than when trained with $\mathcal{D}(en,en,ko)$ (84.93). Similarly, for the mE5-base model on the $\mathcal{T}(ko-en)$ task, training with $\mathcal{D}(ko,en,en)$ (73.26) or $\mathcal{D}(ko,en,ko)$ (73.10) outperforms the base model (65.98) or training with $\mathcal{D}(ko,ko,en)$ (68.62). 
These results indicate that aligning \( (Q_L, P_L) \) with the evaluation direction reliably improves performance on the corresponding CLIR task.

\paragraph{Enhancement by Positive-Negative Language Match}
Also, our results demonstrate that when $Q_L\neq P_L$, using negatives in the same language as the positive ($P_L = N_L$) shows the best performance overall.
For instance, with mE5-large, $\mathcal{D}(ko,en,en)$ outperforms $\mathcal{D}(ko,en,ko)$ on OVR (84.56 vs.\ 78.00), and with mgte-base, $\mathcal{D}(en,ko,ko)$ exceeds $\mathcal{D}(en,ko,en)$ (80.81 vs.\ 78.31). These findings suggest that if $Q_L\neq P_L$, using $P_L=N_L$ makes the model learn to distinguish fine-grained semantics in the target language from contrastive learning, thereby boosting CLIR capability.
By contrast, if $Q_L\neq P_L$ and 
% negatives are in the query language ($Q_L{=}N_L$; e.g.,
$P_L\neq N_L$
(e.g., $\mathcal{D}(ko,en,ko)$, $\mathcal{D}(en,ko,en)$), it does acquire some semantic understanding, but it simultaneously learns to prioritize linguistic difference over semantic relevance. This is because the contrastive loss trains the model to pull the positive sample ($P_L$) (which is in a different language) closer to the query ($Q_L$), while pushing the negative sample ($N_L$) (which is in the same language as the query). This enhances the model's sensitivity to language differences but reduces its discernment of subtle target language semantics, thus lowering CLIR performance.

% is most easily minimized by distinguishing the query's language from the positive's, the model's ability to learn semantics within the target language is underdeveloped than Positive-Negatives Match.

% Conversely, if $Q_{L}\neq P_{L} \;\&\; Q_{L}=N_{L}$ (e.g., $\mathcal{D}(koenko)$, $\mathcal{D}(enkoen)$), 
% the model prefers any language differing from the query language regardless of its meanings, due to contrastive learning. This can enhance sensitivity to language differences but reduce discernment of subtle target language semantics, thus lowering CLIR performance

\subsection{Mono-IR Retrieval Results}
The right side of Table~\ref{tab:main1} reports the Mono-IR performance for models trained on all data combinations.
In contrast to the CLIR results, Mono-IR performance generally decreases after training.
A notable observation is the large Mono-IR drop when training with $Q_L = N_L$ and $Q_L \ne P_L$ (e.g., $D(ko,en,ko)$ and $D(en,ko,en)$).
For instance, considering the OVR score for mE5-large, training on $D(ko,en,ko)$ (83.61) and $D(en,ko,en)$ (85.12) leads to substantial degradation compared to the base model (88.89).
In contrast, models trained on $D(ko,en,en)$ or $D(en,ko,ko)$, where $Q_L \ne P_L$ and $P_L = N_L$, experience only minor decreases, or even show improvements in the case of mgte-base.
The larger degradation under $Q_L = N_L$ and $Q_L \ne P_L$ can be explained by the language configuration of the negatives.
In these settings, each query is paired with a cross-lingual positive, while its hard negatives are written in the same language as the query.
As a result, the contrastive objective pulls the cross-lingual positive toward the query while repeatedly pushing away same-language negatives.
This can disrupt the same-language semantic neighborhoods that are important for Mono-IR, leading to a larger performance drop.
In other words, the model may learn a language-level separation signal that competes with semantic alignment.

\subsection{Retrieval Results for Merged Models}
Our preceding experiments reveal that optimizing for CLIR performance can lead to a degradation in Mono-IR capabilities, creating a performance trade-off. 
To address this issue, we incorporate a model merging strategy. Inspired by Model Soup~\citep{wortsman2022model}, which is used to resolve task-specific conflicts~\citep{lee2025gemini, zhang2025qwen3}, we investigate whether simple weight-averaged model merging~\citep{matena2022merging} can mitigate such trade-offs.
We merge the base and fine-tuned parameters with a fixed equal ratio, $\theta_{\text{merged}} = 0.5\,\theta_{\text{base}} + 0.5\,\theta_{\text{ft}}$.
Since the data combinations that improved CLIR performance for the mgte-base model did not degrade its Mono-IR performance, we only use the remaining three models in this merging experiment.

\definecolor{mypastelblue}{RGB}{230, 240, 250}

\begin{table}[!ht]
\centering
% \small
\resizebox{1\linewidth}{!}{%
\begin{tabular}{l|cc|cc} % DATA 열을 왼쪽 정렬(l)로 변경
\toprule[1.2pt]
\multicolumn{1}{c|}{\multirow{2}{*}{\textbf{MODEL}}} & \multicolumn{2}{c|}{\textbf{CROSS-AVG}} & \multicolumn{2}{c}{\textbf{MONO-AVG}} \\ % DATA 헤더를 중앙 정렬
\cmidrule(lr){2-3}\cmidrule(lr){4-5}
& en-ko & ko-en & ko-ko & en-en \\
\midrule[1.2pt] % 헤더와 첫 번째 모델 그룹 구분선
\multicolumn{5}{c}{\textit{\textbf{bge-m3}}} \\
\midrule[1.2pt]
\rowcolor{metacolorab} base & 85.81 & 80.01 & 86.29 & 89.99 \\
\hspace*{1pt}└ koenen FT & 86.58 & 81.26 & 85.89 & 89.55 \\
\rowcolor{mypastelblue} % "Merged" 행에 파스텔 블루 적용
\hspace*{1pt}└─ Merged (base+ koenen FT) & 86.96 & 81.21 & 86.49 & 90.08 \\
\hspace*{1pt}└ enkoko FT & 86.53 & 80.87 & 85.80 & 89.54 \\
\rowcolor{mypastelblue} % "Merged" 행에 파스텔 블루 적용
\hspace*{1pt}└─ Merged (base + enkoko FT) & 86.85 & 80.99 & 86.51 & 90.15 \\
\midrule[1.2pt] % 모델 그룹 간 구분선
\multicolumn{5}{c}{\textit{\textbf{multilingual-e5-large (mE5-large)}}} \\
\midrule[1.2pt]
\rowcolor{metacolorab} base & 87.04 & 76.88 & 87.43 & 90.35 \\
\hspace*{1pt}└ koenen FT & 87.99 & 81.13 & 86.01 & 89.63 \\
\rowcolor{mypastelblue} % "Merged" 행에 파스텔 블루 적용
\hspace*{1pt}└─ Merged (base + koenen FT) & 88.81 & 81.73 & 87.68 & 90.52 \\
\hspace*{1pt}└ enkoko FT & 88.04 & 80.44 & 85.76 & 90.04 \\
\rowcolor{mypastelblue} % "Merged" 행에 파스텔 블루 적용
\hspace*{1pt}└─ Merged (base + enkoko FT) & 89.20 & 81.23 & 87.17 & 90.84 \\
\midrule[1.2pt] % 모델 그룹 간 구분선
\multicolumn{5}{c}{\textit{\textbf{multilingual-e5-base (mE5-base)}}} \\
\midrule[1.2pt]
\rowcolor{metacolorab} Base & 80.11 & 65.98 & 84.62 & 89.50 \\
\hspace*{1pt}└ koenen FT & 84.54 & 73.26 & 82.16 & 88.91 \\
\rowcolor{mypastelblue} % "Merged" 행에 파스텔 블루 적용
\hspace*{1pt}└─ Merged (base + koenen FT) & 84.93 & 72.71 & 84.20 & 89.82 \\
\hspace*{1pt}└ enkoko FT & 85.02 & 72.17 & 80.93 & 89.35 \\
\rowcolor{mypastelblue} % "Merged" 행에 파스텔 블루 적용
\hspace*{1pt}└─ Merged (base + enkoko FT) & 85.66 & 72.16 & 83.70 & 90.00 \\

\bottomrule[1.2pt]
\end{tabular}%

}
\caption{Average performance on CLIR and Mono-IR tasks including merged models. 
% The anchor model for merging is the model that showed the highest score in Mono-Lignual Retreival. 
% Results for all merged models are provided in Appendix~\ref{sec:additional_merging_results}.
}
\label{tab:main2}
\end{table}

\paragraph{Merging Effects on Mono-IR}
Table~\ref{tab:main2} compares the IR performance of the base models, models fine-tuned on $\mathcal{D}(koenen)$ and $\mathcal{D}(enkoko)$ (strong CLIR performers), and the merged models. 
Crucially, the merged models consistently improve Mono-IR performance across both tasks compared to the individually fine-tuned CLIR-optimized models. For example, merging the mE5-large model trained on $\mathcal{D}(koenen)$ with the base model increases Mono-IR score by approximately 1 point on both tasks, in some cases even surpassing the original base model's performance.
\paragraph{Merging Effects on CLIR}
Encouragingly, model merging does not lead to a substantial decline in CLIR performance. For mE5-base, the merged models maintain CLIR performance comparable to the individually fine-tuned models. Moreover, bge-m3 and mE5-large even show slight CLIR improvements after merging.

These findings indicate that weight-averaged model merging with the base model provides a simple and practical way to mitigate the trade-off between CLIR gains and Mono-IR degradation. Moreover, our results provide empirical evidence that model merging can, in some instances, further enhance CLIR performance beyond what is achieved by individually trained specialist models.

\section{Conclusion}
In this study, we conduct an in-depth analysis of how training data language composition impacts CLIR and Mono-IR in multilingual embedding models. We generated linguistically parallel training data and trained models with various language combinations. Through our experiments, we revealed a consistent trade-off: while aligning the languages of positives and negatives substantially enhances CLIR performance, this configuration can concurrently degrade Mono-IR capabilities. To address this trade-off, we demonstrate that simple weight-averaged model merging can partially mitigate Mono-IR degradation while preserving most CLIR gains. Our findings provide valuable insights for future data design and training for robust Cross-Lingual and Mono-Lingual Information Retrieval, and further suggest a methodology applicable to other language pairs.

\section*{Limitations}
This study is subject to several limitations that open avenues for future research. 

First, our investigation primarily focuses on Korean-English CLIR and Mono-IR. Consequently, the generalizability of our findings and the observed effectiveness of the model merging strategy to other language pairs, particularly those with typological or cultural differences, require further validation. 

Second, the synthetic training data, despite efforts to ensure diversity using LLMs and the Korean Decimal Classification, may not fully capture the nuances and breadth of real-world user-generated query-document pairs, especially in highly specialized or niche domains. 

Third, we employ a relatively straightforward Weight-Averaged Model Merging approach. Exploring more sophisticated or task-specific merging methodologies, such as those involving differential layer weighting or the use of Fisher information, might yield different outcomes or superior performance improvements.

Finally, the observed trade-offs and merging effects are contingent on the specific evaluation datasets (Belebele and StrategyQA) used. Different datasets or additional evaluation metrics could reveal different patterns or magnitudes of these effects.

\section*{Ethics Statement}
Synthetic training data in this study was generated via Large Language Models (LLMs) strictly for research on Cross-Lingual and Mono-Lingual Information Retrieval, guided by the Korean Decimal Classification (KDC) to ensure diverse and neutral content. We utilized publicly available models and datasets, promoting transparency and reproducibility. Throughout the research process, including drafting and refining textual content for this paper, an AI assistant was utilized; however, all core research ideas, experimental design, analysis, and final content were directed and validated by the human authors. While acknowledging potential LLM biases, our focus was on the structural properties of synthetic data for training retrieval models, not direct user-facing content generation. Our goal is to improve multilingual information access. The findings aim to contribute to robust and balanced retrieval systems, and we advocate for responsible AI development considering societal impacts.

\section*{Acknowledgments}
This research was supported by Basic Science Research Program through the National Research Foundation of Korea(NRF) funded by the Ministry of Education(NRF-2021R1A6A1A03045425). This work was supported by Institute for Information \& communications Technology Promotion(IITP) grant funded by the Korea government(MSIT) 
(RS-2024-00398115, Research on the reliability and coherence of outcomes produced by Generative AI). This work was supported by Institute for Information \& communications Technology Planning \& Evaluation(IITP) grant funded by the Korea government(MSIT) (No. RS-2022-II220369, (Part 4) Development of AI Technology to support Expert Decision-making that can Explain the Reasons/Grounds for Judgment Results based on Expert Knowledge)

\bibliography{custom}
\clearpage

\appendix

\label{sec:appendix}

% \clearpage
\section{Details for Synthetic Data Generation}
\label{sec:data_details}

\subsection{Prompt Examples}

We provide representative prompts used in our synthetic data generation pipeline.
The pipeline consists of two stages: (1) task brainstorming, which produces diverse retrieval task descriptions conditioned on a KDC topic, and (2) triple generation, which creates bilingual \{query, positive, hard\_negative\} examples for each task.
To support scalable filtering and automatic parsing, we enforce strict output formats: Prompt~1 outputs a Python list of strings, and Prompt~2 outputs a single JSON object.

\paragraph{Prompt Variables}
The prompts are parameterized with placeholders to control diversity and ensure coverage.
In Table~\ref{prompt1} (Prompt~1), \texttt{\{GIVEN\_TOPIC\}} denotes the sampled KDC topic and \texttt{\{NUM\_TASKS\}} controls the number of task descriptions.
In Table~\ref{prompt2} (Prompt~2), \texttt{\{TASK\_DESCRIPTION\}} is a single task description sampled from Prompt~1 outputs.
We further vary query and document attributes via the following placeholders:
\begin{itemize}
    \item \texttt{\{query\_type\}} $\in$ \{extremely long-tail, long-tail, common\}
    \item \texttt{\{query\_length\}} $\in$ \{less than 5 words, 5 to 15 words, at least 15 words\}
    \item \texttt{\{difficulty\}} $\in$ \{high school, college, PhD\}
    \item \texttt{\{clarity\}} $\in$ \{clear, understandable with some effort, ambiguous\}
    \item \texttt{\{num\_words\}} $\in$ \{50, 100, 200, 300, 400, 500\}
\end{itemize}

\paragraph{Prompt 1: Task Brainstorming}
Prompt~1 (Table~\ref{prompt1}) generates a list of retrieval task descriptions in Korean.
Each task specifies what the query represents and what constitutes a relevant document, while discouraging overly narrow tasks to improve coverage.

\begin{table}[h!]
\centering
\small
\renewcommand{\arraystretch}{1.05}
\begin{tabular}{|p{0.97\columnwidth}|}
\hline
{\ttfamily\footnotesize
Brainstorm a list of potentially useful text retrieval tasks related to the given topic. Here are a few examples for your reference:\newline
\ \ - Provided a scientific claim as query, retrieve documents that help verify or refute the claim.\newline
\ \ - Search for documents that answers a FAQ-style query on children's nutrition.\newline
Please adhere to the following guidelines:\newline
\ \ - Specify what the query is, and what the desired documents are.\newline
\ \ - Each retrieval task should cover a wide range of queries, and should not be too specific.\newline
Given topic: \{GIVEN\_TOPIC\}\newline
Your output should always be a python list of strings only, with about \{NUM\_TASKS\} elements, and each element corresponds to a distinct retrieval task in one sentence written in Korean related to the given topic. Do not explain yourself or output anything else. Be creative!
}
\\ \hline
\end{tabular}
\caption{Prompt 1: Task Brainstorming}
\label{prompt1}
\end{table}

\paragraph{Prompt 2: Triple Generation}
Prompt~2 (Table~\ref{prompt2}) generates one training instance in a strict JSON format, including Korean and English versions of each component.
A hard negative is designed to be topically plausible yet incorrect with respect to the task requirement, encouraging the retriever to learn fine-grained relevance beyond superficial overlap.

\begin{table}[h!]
\centering
\small
\renewcommand{\arraystretch}{1.0}
\begin{tabular}{|p{0.97\columnwidth}|}
\hline
{\ttfamily\footnotesize
You have been assigned a retrieval task: \{TASK\_DESCRIPTION\}\newline
Your mission is to write one text retrieval example for this task in JSON format.\newline
The JSON object must contain the following keys:\newline
\ \ - "user\_query (kor)": a string, a random user search query specified by the retrieval task.\newline
\ \ - "user\_query (eng)": a string, an English version of "user\_query (kor)"\newline
\ \ - "positive\_document (kor)": a string, a relevant document for the user query.\newline
\ \ - "positive\_document (eng)": a string, an English version of "positive\_document (kor)"\newline
\ \ - "hard\_negative\_document (kor)": a string, a hard negative document that only appears relevant to the query.\newline
\ \ - "hard\_negative\_document (eng)": a string, an English version of "hard\_negative\_document (kor)"\newline
Please adhere to the following guidelines:\newline
\ \ - The "user\_query" should be \{query\_type\}, \{query\_length\}, \{clarity\}, and diverse in topic.\newline
\ \ - All documents should be at least \{num\_words\} words long.\newline
Your output must always be a JSON object only, do not explain yourself or output anything else. Be creative!
}
\\ \hline
\end{tabular}
\caption{Prompt 2: Triple Generation}
\label{prompt2}
\end{table}

% \paragraph{Post-processing}
% After generation, we parse the outputs and apply lightweight filtering, including format validation and deduplication.

\paragraph{Post-processing}
After generation, we parse the model outputs and apply lightweight filtering.
First, we perform format validation: any sample that cannot be parsed as a JSON object (or does not contain the required keys) is discarded.
We then apply near-deduplication to reduce redundancy in the synthetic training set.
Concretely, we deduplicate based on the English query field (\texttt{user\_query (eng)}) using MinHash LSH~\citep{broder1997resemblance} with a Jaccard-similarity threshold of 0.5 and 128 permutations.
We tokenize each query using the \texttt{FacebookAI/xlm-roberta-large\footnote{\url{FacebookAI/xlm-roberta-large}}} tokenizer, build a MinHash signature over the resulting token set, and insert it into an LSH index.
We retrieve candidate duplicates via LSH queries, and for each detected duplicate pair we remove the later-occurring instance.
To avoid over-pruning across unrelated topics, we first run this MinHash deduplication within each KDC category and then run the same procedure once again globally over the merged set to eliminate remaining cross-category duplicates. The number of instances in the deduplicated dataset is 33k.

% We further mine multiple hard negatives per query to strengthen the training signal.
% Since Prompt~2 provides bilingual fields for each component, we can instantiate controlled Korean--English language compositions by selecting either the Korean or English version for each of \{query, positive, negative\}, enabling systematic analysis in our experiments.

% \begin{table}[t]
% \centering
% \small
% \begin{tabular}{c|ccc}
% \hline
% Variant & Query & Positive & Negative \\
% \hline
% 1 & Kor & Kor & Eng \\
% 2 & Kor & Eng & Kor \\
% 3 & Kor & Eng & Eng \\
% 4 & Eng & Eng & Kor \\
% 5 & Eng & Kor & Eng \\
% 6 & Eng & Kor & Kor \\
% \hline
% \end{tabular}
% \caption{Six Korean--English cross-lingual variants used in training (excluding fully monolingual cases).}
% \label{tab:variants}
% \end{table}

% \newpage

% \subsection{Details for Hard Negative Mining}
% \label{appen_hn_mining}
% In this section, we describe our hard negative mining procedure for constructing an effective training dataset based on the generated synthetic data.

% We precompute embeddings for all English positive documents (\texttt{positive\_document (eng)}) using the NV-Embed-v2~\citep{moreira2024nv} model. Hard negative mining is then performed for each English query based on these embeddings. For each query, documents are ranked in descending order of similarity, and candidates after the top 50th rank are considered.
% Using this process, we mine 5 English hard negatives per query. Finally, by adding the originally generated synthetic hard negative, we obtain a total of 6 hard negatives for each query.
\subsection{Details for Hard Negative Mining}
\label{appen_hn_mining}
In this section, we describe our hard negative mining procedure for constructing an effective training dataset based on the generated synthetic data.

Our synthetic triples are generated to be fully parallel across languages, i.e., each instance provides aligned query/positive/negative texts in both English and Korean that correspond to the same underlying semantics. 
This one-to-one parallel structure allows us to mine hard negatives in one language and reuse them for other language variants without changing the intended negative meaning. 
Accordingly, we perform hard negative mining on the English side and transfer the mined negatives to other variants via their parallel counterparts.

Concretely, we precompute embeddings for all English positive documents (\texttt{positive\_document (eng)}) using NV-Embed-v2~\citep{moreira2024nv}, and mine hard negatives for each English query based on these embeddings. 
We intentionally use NV-Embed-v2 (instead of a multilingual retriever) because hard negative mining benefits from a model that is strong at fine-grained semantic discrimination in the mining language; in our experience, a high-performing English retriever yields sharper similarity rankings and thus more challenging negatives. 
For each query, we rank documents by similarity, ignore the top-50 retrieved candidates to reduce potential false negatives, and then sample 5 English hard negatives per query from the remaining candidates. 
Finally, by adding the originally generated synthetic hard negative, we obtain a total of 6 hard negatives for each query. 
For variants involving Korean queries or Korean negatives, we directly use the Korean counterparts of these mined negatives provided by the parallel synthetic generation, so that language variants share the same underlying negative semantics while differing only in language.

\subsection{Final Training Dataset}
Finally, we instantiate language-composed training data by selecting either the Korean or English translation for each of \{query, positive, hard\_negative\}, resulting in six Korean--English cross-lingual variants that cover all language configurations.

\begin{table}[h!]
\centering
\small
\begin{tabular}{c|c|ccc}
\hline
\# of Data & Variant & Query & Positive & Negatives(6) \\
\hline
\multirow{6}{*}{33k}
  & 1 & Kor & Kor & Eng \\
  & 2 & Kor & Eng & Kor \\
  & 3 & Kor & Eng & Eng \\
  & 4 & Eng & Eng & Kor \\
  & 5 & Eng & Kor & Eng \\
  & 6 & Eng & Kor & Kor \\
\hline
\end{tabular}
\caption{Six Korean--English cross-lingual variants used in training (excluding fully monolingual cases).}
\label{tab:variants}
\end{table}

We provide an example of one final training data composition, $\mathcal{D}(\textit{ko,en,en})$, in Table~\ref{tab:train_example}.
The example illustrates a cross-lingual training triple where an ``Anchor'' query in Korean is paired with a ``positive'' document in English that directly answers the query.
The ``negative'' documents are topically related to the anchor (e.g., French learning, grammar, or dialects for a query about French pronunciation rules), yet they do not satisfy the information need, making them challenging distractors during training.
This structure is designed to train a system for cross-lingual information retrieval, emphasizing semantic understanding.

\subsection{Human Quality Assessment}
\label{appen:human_eval}

To assess the quality of our synthetic training dataset, we conduct a manual human evaluation on a random sample of generated instances.
We sample 100 training instances in total, consisting of 50 English-only instances ($\mathcal{D}(\textit{en,en,en})$) and 50 Korean-only instances ($\mathcal{D}(\textit{ko,ko,ko})$).
Four expert annotators, two native English speakers and two native Korean speakers, evaluate each sample using a 3-point scale (0--2); for each language group, one annotator is an M.S. student and the other is a Ph.D. student in Computer Science. The criteria are as follows:
\begin{itemize}
    \item Grammatical Fluency (0--2): whether the query is natural and clear (2 = native-level, 0 = unintelligible).
    \item Positive Relevance (0--2): whether the positive document contains an answer to the query (2 = explicit answer, 0 = irrelevant).
    \item Hard Negative Validity (0--2): whether the hard negative is topically related but off-target (2 = valid hard negative, 0 = false negative / actually correct).
\end{itemize}

Table~\ref{tab:human_eval} summarizes the results (mean $\pm$ std), where we additionally report the percentage of the maximum possible score in parentheses.
Overall, the evaluation yields consistently high scores across all criteria.
In particular, positive documents are almost always judged to directly answer the query (approximately 99\% of the maximum score), and hard negatives are largely valid while remaining non-answering distractors (approximately 80--84\% of the maximum score).
Despite the inherent subjectivity of linguistic judgments, these results indicate strong annotator agreement on the quality of the generated corpus, supporting its suitability for retrieval model fine-tuning.

\begin{table}[ht]
\centering
% \small
\resizebox{\linewidth}{!}{%
\begin{tabular}{l|cc}
\hline
\textbf{Metric} & $\mathcal{D}(\textit{en,en,en})$ & $\mathcal{D}(\textit{ko,ko,ko})$ \\
\hline
Fluency      & $1.82 \pm 0.39$ (91\%) & $1.82 \pm 0.39$ (91\%) \\
Pos.\ Rel.   & $1.98 \pm 0.14$ (99\%) & $1.97 \pm 0.14$ (99\%) \\
HN Validity  & $1.68 \pm 0.51$ (84\%) & $1.57 \pm 0.54$ (80\%) \\
\hline
\end{tabular}%
}
\caption{Human quality assessment of the synthetic training corpus (mean $\pm$ std over a 0--2 scale). Percentages denote the ratio to the maximum possible score.}
\label{tab:human_eval}
\end{table}

\begin{table*}[t]
\centering
\begin{tabular}{|p{2cm}|p{14cm}|}
\hline
\textbf{Type} & \textbf{Content} \\
\hline
Anchor & 프랑스어의 발음 규칙은 어떻게 되나요? \textit{(What are the pronunciation rules in French?)} \\
\hline
Positive & The pronunciation rules of French are relatively clear and systematic. The French alphabet consists of 26 letters, each with its own unique pronunciation rules. For instance, 'e' is often silent when it appears at the end of a word or is unstressed. Additionally, 'h' is generally not pronounced, and 'c' is pronounced as 's' when followed by 'e', 'i', or 'y'. These rules help French learners to understand pronunciation more easily. \\
\hline
Negative 1 & French is a Romance language primarily spoken in France, Belgium, Switzerland, and Canada. It is spoken by approximately 270 million people and is one of the six official languages of the United Nations. French originated from Latin and has evolved over time into modern French. The language is known for its distinct pronunciation, particularly the unique pronunciation of the letter 'r' compared to other languages. Learning French offers an opportunity to deepen understanding of the language's history and culture. \\
\hline
Negative 2 & For practicing French pronunciation, it is important to focus on particularly difficult or commonly mispronounced sounds. Since pronunciation is very important in French, constant repetition and practice are necessary to get it right. The 'r' sound and nasal vowels are particularly challenging for many learners, and special exercises may be required for these. \\
\hline
Negative 3 & French verb conjugation has a complex structure. Generally, verbs are divided into three main groups: regular -er verbs, -ir verbs, and irregular verbs. Each group follows different conjugation rules depending on various tenses such as present tense, past tense, and future tense. For instance, the present form of -er verbs changes according to the endings, and compound tenses are formed by using the appropriate form of 'avoir' or 'être' along with the past participle. In compound past tense, 'avoir' is mainly used, but for verbs of movement or state change, 'être' is used as an exception. These rules form the basis of French verb conjugation, and understanding the irregularities of each verb is crucial. \\
\hline
Negative 4 & The most important aspect of starting to learn French is understanding the basic grammar rules. Nouns, adjectives, and verbs in French change form according to gender and number, and mastering these rules is essential. Additionally, French pronunciation is quite different from English, so practicing pronunciation is also necessary. To learn French, it is advisable to use a variety of learning materials. For example, reading French novels or news articles can be beneficial. Hiring an online tutor or participating in language exchange programs can also be a great way to practice French conversation. Finally, consistency is key in learning French, so it's important to continue with daily study, even if it's just a little each day. \\
\hline
Negative 5 & French spelling often does not match its pronunciation, which can confuse learners. For instance, 'beaucoup' is pronounced as 'boku,' but its spelling does not suggest this. These discrepancies are related to the historical development of the French language. Additionally, there are various accents and pronunciations in French that can vary by region. Specifically, the French 'r' is different from the English 'r,' which can be challenging for learners to grasp initially. These mismatches between spelling and pronunciation are significant challenges to overcome when learning French. \\
\hline
Negative 6 & Understanding the important rules of French grammar is fundamental to language learning. Key elements of French grammar include the agreement of gender and number in nouns and adjectives, verb tense and conjugation, pronoun placement, and the use of articles. In French, nouns and adjectives are divided into masculine and feminine forms, and their forms change depending on whether they are singular or plural. Verbs change according to the subject and tense, and it is particularly important to learn the conjugation of regular and irregular verbs. Furthermore, pronouns are used according to their position and role in a sentence. Lastly, articles change according to the gender and number of the noun, and understanding the difference between definite and indefinite articles is essential. \\
\hline
\end{tabular}
\caption{Training Data Example of $\mathcal{D}(\textit{koenen})$}
\label{tab:train_example}
\end{table*}

\subsection{Experimental Setup Details}
\label{appen_exp_setup}
All models were trained for a total of one epoch with a batch size of 512, employing a linear learning rate scheduler with a warm-up ratio of 0.1. We used the AdamW optimizer~\citep{loshchilov2017decoupled} (with parameters $\beta_{1}=0.9$, $\beta_{2}=0.99$, and weight decay=0.01), and adopted bfloat16 mixed precision for computational efficiency. The learning rate for all models is fixed to 2e-5.

We used 2 NVIDIA A100 GPUs, each with 80GB of memory capacity, along with AMD EPYC 7513 processors featuring 32 cores, to train models. 
Additionally, to increase the batch size within limited GPU resources, we adopt CachedGISTEmbedLoss from sentence-transformers~\citep{reimers2019sentence}, which combines the GISTEmbedLoss with the gradient cache technique~\cite{gao2021scaling}. We use bge-m3 as a guide model for the GISTEmbedLoss. For evaluation, we employed a single A100 GPU.

% \subsection{Synthetic Data Example}
% \label{appen_synthetic_data}
% % We provide examples of Generated Synthetic Data. 
% Table~\ref{tab:synthetic_example} shows an example of well-constructed synthetic data. The user query, "Changes in French presidential election system," is precise, clearly defining the scope of interest. The positive document demonstrates high relevance by directly addressing this query, outlining specific historical changes to the French presidential election system, including the introduction of direct elections and the reduction of the presidential term. The hard negative document is strategically crafted: it discusses the French local election system. This document shares keywords and the general theme of "French elections," making it a challenging distractor. However, its focus on local elections, rather than the presidential system specified in the query, makes it a true negative. This distinction is crucial for teaching models to discern nuanced differences in query intent beyond superficial keyword overlap.
% \input{tables/Appendix_synthetic_example}

\clearpage
\section{Evaluation Benchmark}
\label{sec:appen_benchmark}
In this work, we leverage multilingual Question Answering (QA) datasets with parallel constructions, repurposed as retrieval tasks, to systematically assess CLIR performance in all directions. Since these datasets are originally designed for QA, the corresponding passages serve as exact gold labels within the retrieval framework. Consequently, datasets developed for QA (query and its corresponding passage for answering) are widely adopted for the evaluation of the retriever in the current literature~\cite{enevoldsen2025mmteb, lee2025gemini, zhang2025qwen3}.

\paragraph{Belebele} Belebele\footnote{\url{https://github.com/embeddings-benchmark/mteb/blob/main/mteb/tasks/retrieval/multilingual/belebele_retrieval.py}} is a high-quality, professionally translated multilingual QA dataset featuring a broad range of language pairs. All translations were conducted by native speakers proficient in both English and target language, thereby capturing both contextual meaning and cultural nuances. Owing to these strengths, Belebele offers diverse and realistic multilingual retrieval scenarios, enabling detailed comparative analyses of retrieval models across different languages.

\paragraph{StrategyQA}
StrategyQA is a crowdsourced dataset of diverse yes/no questions that often require implicit multi-step reasoning.
Ko-StrategyQA\footnote{\url{https://github.com/embeddings-benchmark/mteb/blob/main/mteb/tasks/retrieval/kor/ko_strategy_qa.py}} was created by translating the original StrategyQA questions and evidence paragraphs using DeepL, and is included in the MMTEB benchmark.
In our experiments, we additionally use the original English StrategyQA as a paired counterpart of Ko-StrategyQA to form an English--Korean parallel evaluation setting aligned by construction.

Beyond these, our evaluation was limited by the lack of other retrieval datasets that provide parallel Korean-English pairs.

\section{Robustness to Hard Negative Mining Strategies}
\label{sec:additional_hn_results}
We conduct additional experiments by adopting the hard negative mining strategies proposed in NV-Retriever~\cite{moreira2024nv}. This paper suggests that utilizing absolute 0.05 margin and relative 0.05 margin in the similarity score yields optimal results for hard negative mining. We follow this procedure and repeat the \textit{bge-m3} experiments under both mining strategies. The results are provided in Table~\ref{tab:appen_hnmine_abs} and Table~\ref{tab:appen_hnmine_rel}. 
% These new experiments, conducted with the bge-m3 model using the specified hard negative mining strategies, consistently reinforce the core findings presented in our paper mentioned in Section~\ref{sec:analysis}. 
\paragraph{Results}
Although different mining strategies lead to different absolute performance levels, we observe that the qualitative trends reported in Section~\ref{sec:analysis} remain consistent under the additional experiment. 
% Specifically, the relative comparisons between training configurations are preserved, and the same direction of changes is observed in both overall retrieval metrics and target-language focused metrics.
Specifically, the relative comparisons between training configurations are largely preserved,
and similar trends are observed across both CLIR and Mono-IR settings.

% These results indicate that our main conclusions are robust to (i) the choice of retriever backbone and (ii) the way hard negatives are constructed.

\begin{table*}[!h]
\centering
\small
\resizebox{1.0 \linewidth}{!}{%
\begin{tabular}{c|cc|cc!{\vrule width 1pt}cc|c||cc|cc!{\vrule width 1pt}cc|c}
\toprule[1.2pt]
\multicolumn{15}{c}{\textit{\textbf{bge-m3 with Absolute 0.05 Margin Hard Negative Mining}}}\\
\midrule[1.2pt]
\multirow{4}{*}{\textbf{Data}} &
\multicolumn{7}{c||}{\textbf{Cross-Lingual}} &
\multicolumn{7}{c}{\textbf{Mono-Lingual}} \\
\cmidrule(lr){2-8}\cmidrule(lr){9-15}
& \multicolumn{2}{c|}{\textbf{Belebele}} & \multicolumn{2}{c!{\vrule width 1pt}}{\textbf{StrategyQA}} & \multicolumn{2}{c|}{\textbf{AVG}} & \multirow{2}{*}{\textbf{OVR}} &
\multicolumn{2}{c|}{\textbf{Belebele}} & \multicolumn{2}{c!{\vrule width 1pt}}{\textbf{StrategyQA}} & \multicolumn{2}{c|}{\textbf{AVG}} & \multirow{2}{*}{\textbf{OVR}} \\
\cmidrule(lr){2-3}\cmidrule(lr){4-5}\cmidrule(lr){6-7} \cmidrule(lr){9-10}\cmidrule(lr){11-12}\cmidrule(lr){13-14}
\scalebox{1}{Q-P-N} & en-ko & ko-en & en-ko & ko-en & en-ko & ko-en & & ko-ko & en-en & ko-ko & en-en & ko-ko & en-en & \\
\midrule[1.2pt]
\rowcolor{metacolorab} base & 90.37 & 88.36 & 81.24 & 71.65 & 85.81 & 80.01 & 82.91 & 93.16 & 95.55 & 79.41 & 84.42 & 86.29 & 89.99 & \textbf{88.14} \\
% kokoko & \cellcolor{lightred} 90.07 & \cellcolor{lightred} 87.73 & \cellcolor{lightgreen} 81.35 & \cellcolor{lightred} 71.63 & \cellcolor{lightred} 85.71 & \cellcolor{lightred} 79.68 & \cellcolor{lightred} 82.70 & \cellcolor{lightred} 92.71 & \cellcolor{lightred} 95.08 & \cellcolor{lightgreen} 79.51 & \cellcolor{lightgreen} 84.54 & \cellcolor{lightred} 86.11 & \cellcolor{lightred} 89.81 & \cellcolor{lightred} 87.96 \\
koenko & \cellcolor{lightred} 89.67 & \cellcolor{mediumgreen} 89.51 & \cellcolor{mediumred} 79.41 & \cellcolor{mediumgreen} 72.78 & \cellcolor{mediumred} 84.54 & \cellcolor{mediumgreen} \textbf{81.15} & \cellcolor{lightred} 82.84 & \cellcolor{mediumred} 91.90 & \cellcolor{mediumred} 94.42 & \cellcolor{mediumred} 78.08 & \cellcolor{lightred} 83.73 & \cellcolor{mediumred} 84.99 & \cellcolor{mediumred} 89.08 & \cellcolor{strongred} 87.03 \\
kokoen & \cellcolor{lightgreen} 90.68 & \cellcolor{mediumred} 87.18 & \cellcolor{mediumgreen} 82.18 & \cellcolor{mediumred} 70.65 & \cellcolor{lightgreen} 86.43 & \cellcolor{mediumred} 78.92 & \cellcolor{lightred} 82.67 & \cellcolor{lightgreen} 93.16 & \cellcolor{lightred} 95.44 & \cellcolor{lightgreen} 79.66 & \cellcolor{lightred} 84.40 & \cellcolor{lightgreen} 86.41 & \cellcolor{lightred} 89.92 & \cellcolor{lightred} 88.16 \\
koenen & \cellcolor{mediumgreen} 91.49 & \cellcolor{lightgreen} 89.10 & \cellcolor{lightgreen} 81.71 & \cellcolor{mediumgreen} 72.73 & \cellcolor{lightgreen} 86.60 & \cellcolor{lightgreen} 80.92 & \cellcolor{mediumgreen} 83.76 & \cellcolor{lightred} 92.42 & \cellcolor{mediumred} 94.78 & \cellcolor{lightred} 79.36 & \cellcolor{lightred} 84.09 & \cellcolor{lightred} 85.89 & \cellcolor{lightred} 89.44 & \cellcolor{mediumred} 87.66 \\
% enenen & \cellcolor{lightgreen} 90.37 & \cellcolor{lightgreen} 88.41 & \cellcolor{lightgreen} 81.53 & \cellcolor{lightred} 71.52 & \cellcolor{lightgreen} 85.95 & \cellcolor{lightred} 79.97 & \cellcolor{lightgreen} 82.96 & \cellcolor{lightred} 92.70 & \cellcolor{lightred} 95.16 & \cellcolor{lightgreen} 79.49 & \cellcolor{lightgreen} 84.53 & \cellcolor{lightred} 86.10 & \cellcolor{lightred} 89.85 & \cellcolor{lightred} 87.97 \\
enenko & \cellcolor{lightred} 89.65 & \cellcolor{lightgreen} 88.60 & \cellcolor{lightred} 80.41 & \cellcolor{lightred} 71.64 & \cellcolor{lightred} 85.03 & \cellcolor{lightgreen} 80.12 & \cellcolor{lightred} 82.58 & \cellcolor{lightred} 93.07 & \cellcolor{lightred} 95.18 & \cellcolor{lightred} 79.32 & \cellcolor{lightgreen} 84.53 & \cellcolor{lightred} 86.20 & \cellcolor{lightred} 89.86 & \cellcolor{lightred} 88.03 \\
enkoen & \cellcolor{stronggreen} 92.03 & \cellcolor{lightgreen} 88.92 & \cellcolor{lightgreen} 82.06 & \cellcolor{mediumgreen} 72.67 & \cellcolor{stronggreen} \textbf{87.05} & \cellcolor{lightgreen} 80.80 & \cellcolor{stronggreen} \textbf{83.92} & \cellcolor{lightred} 92.34 & \cellcolor{mediumred} 94.87 & \cellcolor{lightgreen} 79.45 & \cellcolor{lightred} 83.86 & \cellcolor{lightred} 85.90 & \cellcolor{mediumred} 89.37 & \cellcolor{mediumred} 87.63 \\
enkoko & \cellcolor{mediumgreen} 91.59 & \cellcolor{mediumgreen} 89.38 & \cellcolor{lightgreen} 81.95 & \cellcolor{mediumgreen} 72.73 & \cellcolor{lightgreen} \underline{86.77} & \cellcolor{mediumgreen} \underline{81.06} & \cellcolor{mediumgreen} \underline{83.91} & \cellcolor{lightred} 92.56 & \cellcolor{lightred} 95.13 & \cellcolor{lightred} 79.06 & \cellcolor{lightred} 84.22 & \cellcolor{lightred} 85.81 & \cellcolor{lightred} 89.68 & \cellcolor{mediumred} 87.74 \\
\bottomrule[1.2pt]
\end{tabular}%
}
\caption{CLIR and Mono-Lingual IR performance for bge-m3 with Absolute 0.05 Margin Hard Negative Mining. Highest scores in \textbf{bold}, second highest \underline{underlined}.}
\label{tab:appen_hnmine_abs}
\end{table*}
\begin{table*}[!h]
\centering
\small
\resizebox{1.0 \linewidth}{!}{%
\begin{tabular}{c|cc|cc!{\vrule width 1pt}cc|c||cc|cc!{\vrule width 1pt}cc|c}
\toprule[1.2pt]
\multicolumn{15}{c}{\textit{\textbf{bge-m3 with Relative 0.05 Margin Hard Negative Mining}}}\\
\midrule[1.2pt]
\multirow{4}{*}{\textbf{Data}} &
\multicolumn{7}{c||}{\textbf{Cross-Lingual}} &
\multicolumn{7}{c}{\textbf{Mono-Lingual}} \\
\cmidrule(lr){2-8}\cmidrule(lr){9-15}
& \multicolumn{2}{c|}{\textbf{Belebele}} & \multicolumn{2}{c!{\vrule width 1pt}}{\textbf{StrategyQA}} & \multicolumn{2}{c|}{\textbf{AVG}} & \multirow{2}{*}{\textbf{OVR}} &
\multicolumn{2}{c|}{\textbf{Belebele}} & \multicolumn{2}{c!{\vrule width 1pt}}{\textbf{StrategyQA}} & \multicolumn{2}{c|}{\textbf{AVG}} & \multirow{2}{*}{\textbf{OVR}} \\
\cmidrule(lr){2-3}\cmidrule(lr){4-5}\cmidrule(lr){6-7} \cmidrule(lr){9-10}\cmidrule(lr){11-12}\cmidrule(lr){13-14}
\scalebox{1}{Q-P-N} & en-ko & ko-en & en-ko & ko-en & en-ko & ko-en & & ko-ko & en-en & ko-ko & en-en & ko-ko & en-en & \\
\midrule[1.2pt]
\rowcolor{metacolorab} base & 90.37 & 88.36 & 81.24 & 71.65 & 85.81 & 80.01 & 82.91 & 93.16 & 95.55 & 79.41 & 84.42 & 86.29 & 89.99 & \textbf{88.14} \\
% kokoko & \cellcolor{lightred} 89.76 & \cellcolor{lightred} 87.74 & \cellcolor{lightgreen} 81.38 & \cellcolor{mediumred} 70.76 & \cellcolor{lightred} 85.57 & \cellcolor{lightred} 79.25 & \cellcolor{lightred} 82.41 & \cellcolor{lightred} 92.63 & \cellcolor{lightred} 95.04 & \cellcolor{lightred} 79.28 & \cellcolor{lightgreen} 84.63 & \cellcolor{lightred} 85.96 & \cellcolor{lightred} 89.84 & \cellcolor{lightred} 87.90 \\
koenko & \cellcolor{lightred} 89.95 & \cellcolor{mediumgreen} 89.52 & \cellcolor{strongred} 79.05 & \cellcolor{mediumgreen} 72.81 & \cellcolor{mediumred} 84.50 & \cellcolor{mediumgreen} 81.17 & \cellcolor{lightred} 82.83 & \cellcolor{mediumred} 92.09 & \cellcolor{lightred} 94.67 & \cellcolor{strongred} 77.51 & \cellcolor{lightred} 83.97 & \cellcolor{mediumred} 84.80 & \cellcolor{lightred} 89.32 & \cellcolor{strongred} 87.06 \\
kokoen & \cellcolor{lightgreen} 90.78 & \cellcolor{mediumred} 87.15 & \cellcolor{lightgreen} 82.09 & \cellcolor{mediumred} 70.75 & \cellcolor{lightgreen} 86.44 & \cellcolor{mediumred} 78.95 & \cellcolor{lightred} 82.69 & \cellcolor{lightred} 93.01 & \cellcolor{lightred} 95.41 & \cellcolor{lightgreen} 79.68 & \cellcolor{lightgreen} 84.53 & \cellcolor{lightgreen} 86.35 & \cellcolor{lightred} 89.97 & \cellcolor{lightred} 88.16 \\
koenen & \cellcolor{mediumgreen} 91.45 & \cellcolor{lightgreen} 89.21 & \cellcolor{lightgreen} 81.63 & \cellcolor{mediumgreen} 72.76 & \cellcolor{lightgreen} 86.54 & \cellcolor{lightgreen} 80.99 & \cellcolor{mediumgreen} \underline{83.76} & \cellcolor{lightred} 92.57 & \cellcolor{lightred} 94.74 & \cellcolor{lightgreen} 79.59 & \cellcolor{lightred} 84.18 & \cellcolor{lightred} 86.08 & \cellcolor{lightred} 89.46 & \cellcolor{mediumred} 87.77 \\
% enenen & \cellcolor{lightred} 90.36 & \cellcolor{lightred} 88.34 & \cellcolor{lightgreen} 81.94 & \cellcolor{lightred} 71.46 & \cellcolor{lightgreen} 86.15 & \cellcolor{lightred} 79.90 & \cellcolor{lightgreen} 83.03 & \cellcolor{lightred} 92.78 & \cellcolor{lightred} 95.12 & \cellcolor{lightgreen} 79.44 & \cellcolor{lightgreen} 84.56 & \cellcolor{lightred} 86.11 & \cellcolor{lightred} 89.84 & \cellcolor{lightred} 87.98 \\
enenko & \cellcolor{lightred} 89.82 & \cellcolor{lightgreen} 88.80 & \cellcolor{lightred} 80.93 & \cellcolor{lightred} 71.28 & \cellcolor{lightred} 85.38 & \cellcolor{lightgreen} 80.04 & \cellcolor{lightred} 82.71 & \cellcolor{lightgreen} 93.40 & \cellcolor{lightred} 95.29 & \cellcolor{lightgreen} 79.43 & \cellcolor{lightgreen} 84.80 & \cellcolor{lightgreen} \underline{86.42} & \cellcolor{lightgreen} \textbf{90.05} & \underline{88.23} \\
enkoen & \cellcolor{mediumgreen} 91.86 & \cellcolor{lightgreen} 88.77 & \cellcolor{lightgreen} 82.03 & \cellcolor{mediumgreen} 72.30 & \cellcolor{mediumgreen} 86.95 & \cellcolor{lightgreen} 80.54 & \cellcolor{stronggreen} \textbf{83.74} & \cellcolor{mediumred} 92.19 & \cellcolor{lightred} 94.77 & \cellcolor{mediumred} 79.19 & \cellcolor{lightred} 83.73 & \cellcolor{lightred} 85.69 & \cellcolor{mediumred} 89.25 & \cellcolor{mediumred} 87.47 \\
enkoko & \cellcolor{mediumgreen} 91.64 & \cellcolor{mediumgreen} 89.61 & \cellcolor{lightgreen} 81.77 & \cellcolor{mediumgreen} 72.77 & \cellcolor{lightgreen} \textbf{86.71} & \cellcolor{mediumgreen} \textbf{81.19} & \cellcolor{mediumgreen} 83.95 & \cellcolor{lightred} 92.74 & \cellcolor{lightred} 94.95 & \cellcolor{lightred} 79.28 & \cellcolor{lightred} 84.12 & \cellcolor{lightred} 86.01 & \cellcolor{lightred} 89.54 & \cellcolor{mediumred} 87.77 \\
\bottomrule[1.2pt]
\end{tabular}%
}
\caption{CLIR and Mono-Lingual IR performance for bge-m3 with Relative 0.05 Margin Hard Negative Mining. Highest scores in \textbf{bold}, second highest \underline{underlined}.}
\label{tab:appen_hnmine_rel}
\end{table*}

\end{document}